\documentclass[aip,jcp,reprint]{revtex4-1}
\usepackage{bm}
\usepackage{epsfig}
\usepackage{amsmath}
\usepackage{amssymb}
\begin{document}

\newcommand{\kT}{k_\mathrm{B} T}
\newcommand{\tw}{t_\mathrm{w}} 
\newcommand{\QQ}{\mathcal{N}_Q}

\title{Controlling crystal self-assembly using a real-time feedback scheme}

\author{Daphne Klotsa}
\affiliation{Department of Chemical Engineering, University of Michigan, 2300 Hayward St., Ann Arbor MI48109-2136, U.S.A.}
\affiliation{Department of Physics, University of Bath, Bath BA2 7AY, United Kingdom}
\author{Robert L. Jack}
\affiliation{Department of Physics, University of Bath, Bath BA2 7AY, United Kingdom}

\begin{abstract}
We simulate crystallisation of hard spheres with short-ranged attractive potentials, as a model self-assembling system.
Using measurements of correlation and response functions, we develop a method whereby the interaction
parameters between the particles are automatically tuned during the assembly process, in order to obtain high-quality crystals and avoid kinetic traps.  
The method we use is independent of the details of the interaction
potential and of the structure of the final crystal -- we propose that it can be applied to a wide range of self-assembling systems.
\end{abstract}

\maketitle

\newcommand{\eb}{\varepsilon_\mathrm{b}}
\newcommand{\kB}{k_\mathrm{B}}

\section{introduction}

In self-assembly~\cite{white02review,sol07}, simple components like colloids and biomolecules 
come together spontaneously, forming ordered structures such as viral capsids~\cite{hagan06,rap08}, 
crystals~\cite{leunissen05,Hynn2007,sear07,nyk08,chen11,romano11,klotsa11}, or DNA origami~\cite{rothemund06}.  
Recent developments in self-assembly include the experimental
synthesis of particles with specific controllable interactions~\cite{nyk08,chen11,pawar09,sac10,kraft12}, 
as well as theoretical and computational demonstrations of the ordered 
phases that such particles can form~\cite{wilber09,akbari09,romano11,martinez11}.
However, even when ordered states are stable, appearance of disordered aggregates 
often frustrates the dynamical self-assembly process.  
As a result, effective assembly typically involves microscopic reversibility:
if bonds are both made and broken during self-assembly then defects are annealed naturally, producing an ordered final 
state~\cite{white02,hagan06,rap08,whitelam09,grant11,klotsa11}.
The twin requirements of a stable ordered structure and the reversibility of bonding usually
mean that assembly is effective only when interaction parameters are tuned to lie within
 a narrow `optimal' range~\cite{hagan06,whitelam09,klotsa11,jankowski11,grant11}.

In some experiments and in simulations, interactions between particles can be manipulated during the assembly process,
in order to optimize assembly conditions and facilitate the kinetics.
For example, a slow cooling process
is used in hierarchical assembly of DNA origami~\cite{rothemund06}, and light exposure in repeated pulses may allow control of nanoparticle
aggregation~\cite{klajn07,jha12}. However, such protocols have, so far, been largely empirical.  
Given a desired ordered structure and a particle with controllable interactions, 
it is far from clear how these interactions should change in time in order to achieve the best self-assembly.  One might expect
that gradually increasing the interaction strength would allow the product structure to form smoothly and reversibly, as in 
Ref~\onlinecite{rothemund06}. But there is little evidence that this scheme is optimal for assembling ordered structures
quickly and reliably. 

In this paper, we develop a method for {\it automatically} choosing protocols (series of steps) by which interactions between particles
should be manipulated in order to achieve the best assembled products in the most efficient way. We do this by measuring the reversibility 
of particle bonding, as assembly is taking place.  The resulting cycle of measurements
and changes in interaction strength forms an elementary feedback loop. 
Using computer simulations, we show that (i)~our feedback scheme quickly 
locates interaction parameters for which assembly is effective, 
and (ii)~it forms higher-quality crystals than assembly with
fixed (time-independent) interactions.
We also find that the protocols that give best assembly are \emph{not} gradual cooling schemes, and we discuss the reasons
for this observation.
The feedback scheme does not rely in any way
on the structure of the ordered (crystal) state -- we expect 
that it can be applied to other self-assembly processes with minimal modification.
Our results provide a proof of concept for this method, which is efficient in automatically finding effective assembly protocols, 
and avoiding metastable disordered states. 

Currently, this method is best-suited to implementation in computer simulation,
where interactions between particles can easily be controlled, and measurements of reversibility are 
relatively simple to make.  
We emphasise that our computer simulations give dynamically-realistic descriptions of the self-assembly process, so if a given set of
time-dependent interactions is effective in simulation, we would expect them to be effective in experiment too.  


\section{Model}

We consider crystallisation of hard spheres with short-ranged attractive isotropic interactions, as a model self-assembly process.  
This system might represent a suspension of colloidal particles interacting through depletion forces~\cite{whitelam09,fortini08,klotsa11},
or a solution of protein molecules used in protein crystallisation~\cite{frenkel97,rosenbaum96,rosenbaum99,sear07}.  
%

We take $N=10,000$ spherical particles with hard cores of diameter $\sigma$, at volume fraction $\phi=0.04$ and temperature $T$.
The particles interact through a pair
potential $U(r)$ which is a square well of depth $\eb$ and range $\xi=0.1\sigma$.  
Phase diagrams for such systems have been calculated (for example) by Liu~\emph{et~al.}~\cite{liu05}, showing a fluid-crystal phase coexistence regime that contains a metastable fluid-fluid critical point.  
The square well potential that we use is 
a coarse-grained representation of attractive interactions: in colloidal suspensions or protein solutions, it accounts for 
the effects of depletion interactions, hydrophobic association and/or screened electrostatic charges.  
The model is therefore somewhat schematic, but
we build on the observation~\cite{rosenbaum96,rosenbaum99,noro2000} that for systems with short-ranged attractions, 
the shape of the potential $U(r)$ is relatively unimportant, with the behaviour being largely dictated by the (reduced) second virial coefficient.
Thus we expect our main findings to depend only weakly on the shape of this potential~\cite{noro2000,liu05}.

To describe the motion of particles dispersed in a solvent, we use overdamped Langevin dynamics.  Particles have
(bare) diffusion constant $D_0$ and we define a time unit $\tau_0=\sigma^2/D_0$, of the order of a Brownian time.
Our model therefore neglects hydrodynamic effects arising from the solvent, and may
underestimate the rates of diffusion for large clusters of colloidal particles~\cite{spaeth11}.  
However, Langevin dynamics do capture the essential physical processes at work in 
self-assembly~\cite{hagan06,whitelam09,wilber09,romano11,klotsa11}: reversible bonding, 
nucleation of ordered phases, the possibility for kinetic trapping, and Ostwald ripening.  We therefore
use this method for computational convenience and for ease of comparison with other studies~\cite{whitelam09,klotsa11}.
We do not expect our neglect of hydrodynamic effects to have strong effects on the competition between reversible bonding and kinetic trapping, 
which will be our main focus in what follows.  We return to possible hydrodynamic effects in Sec.~\ref{sec:mech} below.

To simulate the Langevin dynamics of the system, we use
single particle Monte Carlo (MC) moves.  
Full details of the computational scheme are given in in Ref.~\onlinecite{klotsa11}: we note that the time unit used in this work
is $\tau_0 \approx 270$ MC sweeps. 
As discussed in Refs.~\onlinecite{whitelam09,whitelam11-molsim,sanz10-mc-bd}, single particle MC moves 
give an accurate representation of the Langevin dynamics, as long as the step size is sufficiently small.
If the step size is too large, it is possible that rejected MC moves act to reduce the
diffusion of large clusters, but the results of Ref.~\onlinecite{whitelam09} show that this has little effect on the assembly pathway
or the final product, in a very similar crystallising system.
The initial conditions of all simulations are taken from equilibrated hard sphere systems at the relevant volume fraction.

In a preceding paper [\onlinecite{klotsa11}] we reported simulations of the model system described here, 
with fixed (time-independent) interaction parameters.
We found that for times up to $10^4\tau_0$, appreciable crystallization takes place only for $2.3<\eb/T<3.0$, 
with $\eb/T=2.5$ the optimum value.
For $\eb/T<2.0$ the system is in the stable fluid phase and does not assemble. 
For $\eb/T>4.0$ the system makes bonds that are too strong and
gets kinetically trapped (see also the phase diagram sketched in Fig.~1 of Ref.~\onlinecite{klotsa11}). 

Here, we perform two kinds of computer simulation: (i)~`No-Feed' protocol. The interactions are fixed and time-independent 
(as in Ref.~\onlinecite{klotsa11}). We focus on bond strengths close to the optimal value $\eb/T$=2.5. 
(ii)~`With-Feed' protocol. The interactions change in time according
to the feedback scheme. The initial bond strength $\eb^\mathrm{init}$ is a free parameter: 
the aim of the feedback scheme is that it should tune the system into a good-assembly regime,
independently of $\eb^\mathrm{init}$. We present three representative cases,
$\eb^\mathrm{init}/T=1.0,2.5,7.0$.  If these bond strengths were used in `No-feed' simulations, they would lead to no assembly, 
near-optimal assembly, and kinetic trapping, respectively.  

\section{Feedback scheme}

\subsection{Fluctuation-dissipation theorem}
\label{subsec:fdt}

Previous studies on glasses~\cite{crisanti03,kurchan05,baiesi09}, 
self-assembling systems~\cite{jack07,klotsa11,grant11pre} and gels~\cite{russo10} have used  
out of equilibrium correlation and response functions to characterise dynamical behaviour~\cite{baiesi09,seifert10}.
Our hypothesis~\cite{jack07} is that by focussing on the dynamics of an evolving system, without knowledge of its structure, we can quantify how reversible
the system is and therefore predict whether it is prone to get kinetically trapped or not. 

In practice, we perturb the dynamics of the system, writing the energy as $E=\frac12 \sum_p (1+h_p)E_p$ where
$E_p = \sum_{p'} U(|\bm{r}_p-\bm{r}_{p'}|)$ is the energy of particle $p$ and $h_p$ is a perturbing field applied to that particle.
The total energy in the presence of the perturbation is therefore $E = \sum_{pp'} ( 1 + \frac12(h_p + h_{p'}) ) U(|\bm{r}_p-\bm{r}_{p'}|)$, and we emphasise that the dynamics obeys detailed balance with respect to this energy function whenever the perturbation is imposed.
The perturbation is switched on at time $t_w$. The energy autocorrelation function and the response function for that perturbation are
(respectively)
\begin{align}
C(t,\tw) & = \langle E_p(t) E_p(\tw) \rangle - \langle E_p(t) \rangle \langle E_p(\tw) \rangle \label{equ:CC} \\
\chi(t,\tw) & =\frac{T}{2}  \frac{\partial}{\partial h_p} \langle E_p(t) \label{equ:chi} \rangle .
\end{align}

We make use of the fact
that \emph{at equilibrium} the system is by definition reversible in time and the relation between the correlation 
and response functions is given by the fluctuation-dissipation theorem (FDT)
$
\chi^\mathrm{eqm}(t,\tw) = C^\mathrm{eqm}(t,t) - C^\mathrm{eqm}(t,\tw)
$. 
(For the specific observables considered here, a proof of this FDT was given in Ref.~\onlinecite{klotsa11}.)
In an assembling system, we can therefore evaluate how much the relation between $C(t,t_w)$ and $\chi(t,t_w)$  
deviates from the equilibrium FDT relation. 
We find three broad classes of behaviour: (i) No deviation from FDT. In this case 
the system is truly reversible and already at equilibrium. (ii) Large deviation. The system is highly irreversible and thus 
likely to get kinetically trapped. (iii) Small but finite deviation. The system is reversible at short timescales and becomes 
irreversible at longer timescales.  Here, the difference between `small' and `large' deviations may be calculated from dimensionless
parameters that lie between $0$ and $1$: see below.
We have shown that the case (iii) above is typically correlated with optimal self-assembly~\cite{jack07,klotsa11,grant11pre}.
The reason is that small deviations from FDT behaviour
indicate the right balance between the microscopically reversible particle bonding and
the macroscopically irreversible self-assembly~\cite{grant11pre} -- which are the two requirements for effective assembly.

In the present paper, the feedback scheme is used to evaluate which of the three options best characterises the dynamics
of the system under investigation, and to dynamically tune the interaction strength between the particles 
accordingly, until the dynamics is consistent with case (iii).

\begin{figure}
\includegraphics{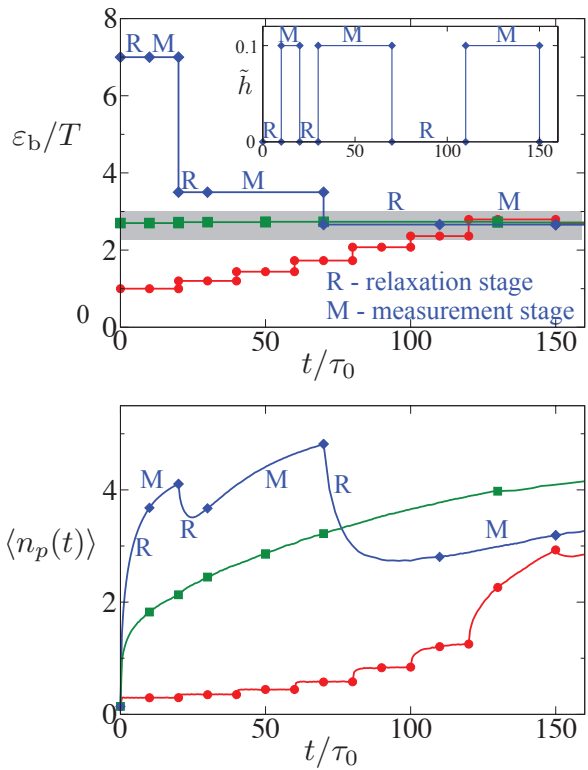}
\caption{Operation of the feedback loop on short time scales for three `With-Feed'  
simulations with different initial bond strengths $\eb^\mathrm{init}$.
(Top)~Time-dependence of the bond strength $\eb$.  Measurement and relaxation stages are indicated for the run
with $\eb^\mathrm{init}/T=7.0$.
The shaded region indicates where `No-Feed' simulations found appreciable crystallisation.
The inset shows the time-dependence of the perturbation strength $\tilde h$ for the run with $\eb^\mathrm{init}/T=7.0$.
(Bottom)~Average number of bonds per particle $\langle n_p(t) \rangle = \frac{\langle E_p(t)\rangle}{(-\varepsilon_\mathrm{b})}$.
}\label{fig:feed}
\end{figure}

\subsection{Implementation of feedback scheme}

The feedback scheme  consists of alternating `measurement' and `relaxation' stages, as shown in Fig.~\ref{fig:feed}.
We describe the main points here:
a full description of the implementation of the scheme is given in the Appendix.
In measurement stages, we perturb the particles with fields $h_p$ (inset to Fig.\ref{fig:feed}) and we measure correlation and response functions.  
The strength of the perturbation is given by a dimensionless parameter $\tilde h$: see Appendix.
At the end of each measurement stage, the bond strength $\eb$ is updated, according to the measurements that have been made.
Each measurement stage is followed by a relaxation stage
during which $h_p=0$; 
after this relaxation stage
the next measurement stage begins. 
      
If the $i$th measurement stage begins at time $t_i$,
it is convenient to normalise correlation and response functions as
$\tilde C(t,t_i) = C(t,t_i)/C(t,t)$ and $\tilde \chi(t,t_i) = \chi(t,t_i)/C(t,t)$.  We define the ratio 
\begin{equation}
\tilde X(t,t_i) = \frac{ \chi(t,t_i) }{ C(t,t) - C(t,t_i) } .
\end{equation}
For a reversible (equilibrated) system, $\tilde X = 1$, in accordance with FDT; for irreversible aggregation we typically find $\tilde X \approx 0$.
This ratio provides a dimensionless measurement of the deviation from reversibility:
we expect that for optimal assembly $\tilde X$ is close but not equal to 1, as in Refs~\onlinecite{jack07,klotsa11,grant11pre}. 
We note that $\tilde X$ differs from the fluctuation-dissipation ratio~\cite{crisanti03} (FDR), which would be an alternative dimensionless measurement.
However, the FDR involves a ratio of derivatives of $\chi$ and $C$, while $\tilde X$ is obtained directly from the values of these functions and is therefore easier to calculate in simulation.  Our previous work~\cite{jack07,klotsa11,grant11pre} indicates that both the FDR and $\tilde X$ are well-correlated with the reversibility of self-assembly.  

At the end of each measurement
stage, the feedback scheme updates the bond strength $\eb$, depending on $\tilde X$.  
We use the simple  update rule
\begin{equation}
\eb^\mathrm{new} = \eb^\mathrm{old} \left[ 1 + \frac{\alpha}{\gamma_i}( \tilde X - X_0) \right].
\label{equ:eb-update}
\end{equation}
Here $X_0$ is the target for the response strength $\tilde X$.  We take $X_0=0.8$,
which makes concrete the requirement in Sec.~\ref{subsec:fdt} that deviations from FDT should be ``small but finite''.
Also, $\alpha=1$ determines the sensitivity of the feedback loop, and $\gamma_i$
is a damping parameter.
We take $\gamma_1=1$ for the first measurement phase; at the end of subsequent measurement phases then $\gamma_i$ is increased by $1$ if the response
$\tilde X$ is within a tolerance $\delta X$ of $X_0$.  We take $\delta X=0.1$.  Further information on the choices of the parameters $(X_0,\alpha,\delta X)$
are given in the Appendix.

\begin{figure*}
\includegraphics{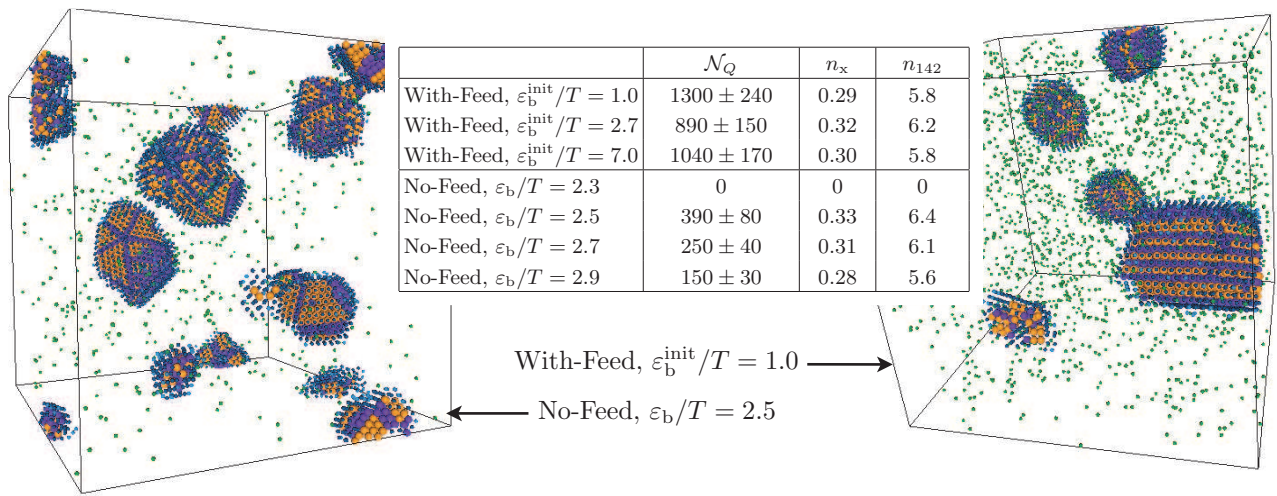} 
%
%
%
%
\caption{Configurations from `No-feed' (left) and `With-feed' (right) simulations at
$t=10^4\tau_0$, and averaged measures of crystallinity at this time.  
In the snapshots, particles in fcc/hcp environments are coloured orange/purple.  
Particles in other environments are rendered at half their actual diameter for visual clarity.
The data in the table are averaged either over 8 independent simulations (`No-Feed') or over the $M=8$ systems associated with
single `With-Feed' simulations (see Appendix).  For $\QQ$, we show the standard errors on these averages, as an indicator of the numerical uncertainty.  For $n_{\rm x}$ and $n_{142}$, these errors are small: no larger than $0.02$ and $0.2$ respectively.  The uncertainties on $\QQ$ are significant but it is clear that `With-feed' simulations form `higher-quality' crystals with larger close-packed domains.
}\label{fig:product}
\end{figure*}

To complete the description of the feedback loop, we specify the durations of measurement and relaxation stages:
that is, the time over which the perturbation $h_p$ is applied, and the time allowed between successive perturbations.
%
These are determined self-consistently: for measurements of reversibility of bonding to be useful, they must be made on appropriate time scales,
and these may not be known \emph{a priori}.
The $i$th measurement stage ends when $t$ is large enough that
either $\tilde C(t,t_i) < C^*$ or $\tilde X(t,t_i) < X^*$, where
$C^*=0.3$ and $X^*=0.6$ are parameters associated with the feedback scheme (see Section~\ref{subsec:duration} below and the Appendix 
for more details).

\section{Results}
\label{sec:results}

In this section we present our simulation results, considering both systems with the feedback scheme on (With-Feed) and with
the feedback scheme off (No-Feed).

\subsection{Feedback loop}
 
The behaviour during the first few stages of the feedback loop is summarised in Fig.~\ref{fig:feed}. 
The first key result of this paper is that for a range of initial bond strengths $\eb^\mathrm{init}$, 
the feedback scheme quickly tunes the interaction strength $\eb/T$ into the (narrow) range where assembly is effective.

More specifically, for $\eb^\mathrm{init}/T=7.0$, the system is vulnerable to kinetic trapping in disordered states, 
but the feedback loop automatically reduces the bond strength to avoid this problem. Similarly for $\eb^\mathrm{init}/T=1.0$, the system
will never assemble if the bond strength is held constant, but the feedback scheme increases $\eb$ in order to achieve assembly. Finally,
for $\eb^\mathrm{init}/T=2.5$ the feedback loop can recognize that the interaction strength is optimal and does not try to
alter the value of $\eb$. This demonstrates
the idea that automated real-time control of interaction parameters can be used to promote effective self-assembly.

\subsection{Crystal structure}


To assess the effectiveness of this scheme,
we analyse the crystals that are assembled within the No-Feed and With-Feed protocols on the relatively long time scale $t=10^4\tau_0$.
Results are summarised in Fig.~\ref{fig:product}.
Results for longer times are qualitatively similar, although there is some slow growth 
in crystalline order as coarsening (Ostwald ripening) takes place.

We measure the crystallinity of the assembled system using both local packing and long-ranged orientational order. 
Local packing is examined using a common neighbour
analysis (CNA)~\cite{hon87}.  The CNA assigns a 4-digit signature to each bonded pair of particles in the system.
Bonds with `1421' or `1422' signatures are characteristic of close-packed crystals: we measure the mean
number of such bonds per particle $n_{142}$, as in Ref.~\onlinecite{klotsa11}.  If every particle is inside a perfect crystal then
$n_{142}=12$.  However, for the finite systems considered in simulation
we always find smaller values of $n_{142}$: this is partly because many particles will be on the surfaces of crystalline
clusters, and partly because the system also contains free monomer particles, and crystals also contain defects.
We also use the CNA to identify particles whose
local environment is consistent with face-centred cubic (fcc) or hexagonal close-packed (hcp) order.  (Fcc particles have 12 bonds with `1422' signatures
while hcp particles have 6 bonds with `1421' signatures and 6 with `1422'.)   If the number of hcp (fcc) particles is $N_\mathrm{hcp}$ ($N_\mathrm{fcc}$) then we
define the fraction of particles in perfectly crystalline environments as $n_\mathrm{x}=(N_\mathrm{hcp}+N_\mathrm{fcc})/N$.  The maximal
possible value for $n_\mathrm{x}$ is unity, but the presence of surface particles, free monomers and crystal defects all act to reduce this.

The table in Fig.~\ref{fig:product} shows that the parameters $n_{142}$ and $n_\mathrm{x}$ are comparable between
the With-Feed protocol (for initial bond strengths $1<\eb^\mathrm{init}/T<7$) and the No-Feed protocol (for the narrower `optimal' range of  bond
strengths $2.5\leq \eb < 2.9$).  Around $30\%$ of particles end the simulation in bulk-crystalline environments.  For these system sizes and densities,
multiple nucleation events occur in all simulations, resulting in several ordered clusters per simulation.  The `maximal yield' of $30\%$ reflects
the fact that particles on the surfaces of the clusters are never classed as crystalline, since they have fewer than 12 bonds.

Fig.~\ref{fig:product} also shows 
snapshots of final configurations where hcp and fcc particles are highlighted: in the example shown, the `With-feed' scheme
has produced a close-packed crystal of around $4400$ particles.
Other small crystalline clusters
are also visible as a result of multiple nucleation.
Other runs under the same conditions yield similar numbers of crystallites of similar
sizes, although there is significant variation due to the stochastic nature of nucleation events.
We emphasise that the large cluster that resulted from the With-Feed protocol is a close-packed crystal,
albeit with random stacking of fcc/hcp planes.    
On the other hand, the `No-Feed' simulations give some clusters where five-fold packing 
defects are apparent, presumably due to growth around a critical nucleus that lacks the symmetry of the crystal.  

To analyse the larger scale order in the assembled crystallites and the presence of packing defects,
we also measure the typical size of crystalline domains in the system using
bond-order parameters.  As in Refs.~\onlinecite{tenwolde95,fortini08}, 
let $\vec{q}_6(p)$ consist of the projection of the bonds of particle $i$ onto the spherical harmonics with $l=6$, 
normalised so that $\vec{q}_6(p)^* \cdot  \vec{q}_6(p)= 1$.  Then, if $\vec{Q}=\sum_{p=1}^N \vec{q}_6(p)$ 
then $\QQ = N^{-1} \langle \vec{Q}^* \cdot \vec{Q}\rangle$ is an estimate
of the typical crystalline domain size.  (To see this, we assume that $\vec{q}_6(p)^*\cdot \vec{q}_6(p')\approx 1$ if particles $p$ and $p'$
belong to the same domain, with $\vec{q}_6(p)^*\cdot \vec{q}_6(p')\approx 0$ otherwise.)  Of course, systems contain a distribution of domain sizes: the quantity $\QQ$ gives equal weight to each particle, so large clusters have the largest contributions to $\QQ$.  Since equal weight is given to each particle, this measure of crystalline domain size is appropriate for measuring the yield of the crystallisation process.

Consistent with the snaphots in Fig.~\ref{fig:product}, the 
numerical results for $\QQ$ in the associated table show that the `With-feed' protocol results in 
significantly larger crystalline domains, compared to its `No-feed' counterpart.
There are differences in crystalline yield between different `With-Feed' runs, coming partly from the choice of initial
bond strength and partly from fluctuations of $C(t,t_i)$ and $\chi(t,t_i)$ in the measurement stages.
However, the trends shown here are robust to these differences.

In evaluating the effectiveness of crystal self-assembly,
we use $\QQ$ as our main `figure-of-merit', since applications of 
self-assembled crystals in photonics or X-ray diffraction both
require well-developed Bragg resonances and hence large ordered domains.
The `With-feed' protocol produces larger values of $\QQ$ than the `No-feed' simulations -- we conclude that 
the feedback scheme does facilitate crystallisation.
This is our second key result, demonstrating explicitly how automated changes in time-dependent interaction
parameters can be used to aid self-assembly.

\section{Discussion}

The effectiveness of the feedback scheme relies on two central assumptions: (i) that correlation and response
measurements can be used to obtain useful information about the reversibility of assembly, and (ii) that tuning
the reversibility of assembly is effective in optimizing the self-assembly process.  Here we discuss these two assumptions
in more detail.

\begin{figure}
\includegraphics[width=7cm]{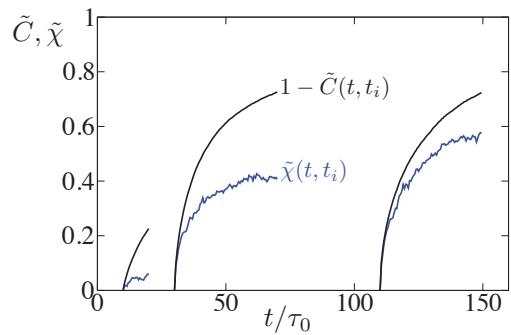}
\caption{
Behaviour of the correlation and response functions for a `With-Feed' simulation with $\eb^\mathrm{init}/T=7.0$. (Other data for the same run was shown in Fig.~\ref{fig:feed}). We show $1 - \tilde{C}(t,t_i)$ and $\tilde\chi(t,t_i)$ for the first
three iterations of the feedback loop.  The responses $\tilde\chi(t,t_i)$ are quite small when compared to $1-\tilde{C}(t,t_i)$, indicating that irreversible bonding is taking place, and the system is vulnerable to kinetic trapping.  As a result, the feedback loop acts to reduce the bond strength $\eb$
(see Fig.~\ref{fig:feed}).
}\label{fig:resp_short}
\end{figure}

\subsection{Duration of measurement and relaxation stages}
\label{subsec:duration}

For effective operation of our feedback scheme,
measurements of reversibility of bonding during assembly must be made on time scales
where significant bond-making and bond-breaking has taken place.  These time scales are not known \emph{a priori}: they depend on the particle
interactions and they also change significantly during the assembly process.  
Hence, within the feedback scheme, we determine the durations of the measurement phases adaptively, using
information contained in the correlation-response measurements themselves.


To illustrate this, consider 
Fig.~\ref{fig:resp_short}, where the behaviour of $\tilde C(t,t_i)$ and $\tilde \chi(t,t_i)$ are shown for one of the `With-Feed' simulations from
Fig.~\ref{fig:feed}.
The functions $1 - \tilde C(t,t_i)$ and $\tilde \chi(t,t_i)$ would coincide at equilibrium: their ratio is equal to $\tilde X$. 
In the first measurement stage, the bonds between
particles are strong and the bonding almost irreversible -- thus, $\tilde \chi$ is small and the ratio $\tilde X$ is much less than its cutoff $X^*$.  As a result, the measurement stage is quickly terminated
and the bond strength reduced.  In the second and third measurement stages, the responses are larger, indicating that
bonding is more reversible.  These stages terminate when $\tilde C$ has fallen to $C^*$, indicating that the system has decorrelated significantly
from its state at time $t_i$: by fixing a value for $C^*$ we ensure that different measurements of 
reversibility can be compared on an equal footing even when the absolute time scales for bonding and unbonding are changing.  
The duration of each relaxation stage is chosen to be equal to the previous measurement stage.  (The only exception is the initial
stage of the feedback loop which is a relaxation stage.  This stage is included because making correlation-response 
measurements immediately after initialization of the system can be sensitive to transient behaviour that depends mostly on the initial
conditions used, and is not necessarily correlated with the effectiveness of self-assembly.
This initial relaxation stage has duration $10\tau_0$ which is sufficient time for initial transients to relax.)

We emphasise that the durations of the 3 measurement stages in Fig.~\ref{fig:resp_short} vary significantly.  On longer time scales,
this variation is even stronger, with the duration tending to increase as the system assembles.  We also investigated schemes where
the durations of measurement stages were fixed \emph{a priori}, but these were less effective.  Allowing these durations to vary within
the scheme does add some complexity, but it does mean that no initial assumption is required as to the relevant time scale for measuring
reversibility.  We argue that this will aid implementation of the scheme in other self-assembling systems where the relevant time scales
may not be known.  Further information about the parameters $C^*,X^*$ associated with the method is given in the Appendix -- we expect
the effectiveness of the feedback loop to depend weakly on these parameters.

\subsection{Time-dependent assembly within feedback scheme}
\label{sec:mech}

\begin{figure}
\includegraphics[width=8cm]{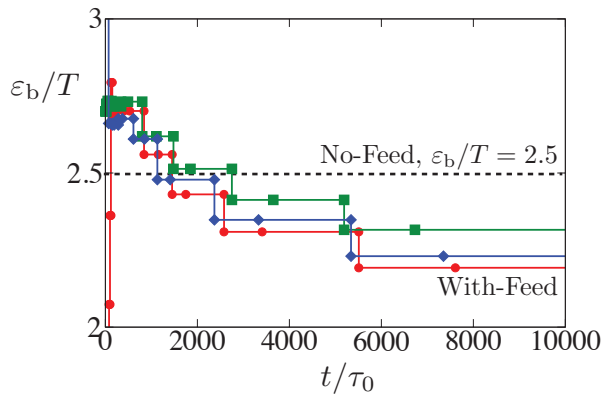}
\caption{
Behaviour of bond strength $\eb$ at long times, for the `With-Feed' simulations of Fig.~\ref{fig:feed}.  The symbols and colors
indicate the values of $\eb^\mathrm{init}$, which are the same as those in Fig.~\ref{fig:feed}.
The dashed line
shows the `No-feed' protocol that gave the best yield.
}\label{fig:beta_long}
\end{figure}

We have emphasised that the feedback loop is designed to exploit the reversibility of bonding in order to promote assembly.
Here we discuss the time-dependent protocols that are selected by the feedback loop, which result in the good assembly shown
in Fig.~\ref{fig:product}.  We show these protocols in Fig.~\ref{fig:beta_long}.
On short time scales, the feedback scheme adjusts the system to have
stronger bonds than the best No-Feed protocol, but on longer time scales the feedback mechanism reduces 
the bond strength.  We note that the With-Feed scheme acts to optimise the reversibility of particle bonding:
it is not obvious \emph{a priori} that this approach of weakening bonds on long times should be effective in optimising
either reversibility or assembly.  In this sense, the `With-feed' scheme is effective in arriving at non-trivial dynamical
approaches for manipulating particle interactions.

Inspection of the assembly trajectories shows that for the bond strengths shown in Fig.~\ref{fig:beta_long}, 
the short-time physics ($t\lesssim 100\tau_0$) is dominated by nucleation, while the behaviour on longer times is that
of Ostwald ripening, with smaller clusters shrinking and larger ones growing.  In this case, the With-Feed protocol leads naturally 
to a scheme where the bonds are initially strong, promoting nucleation; on later time scales the bonds are weaker, which promotes rapid Ostwald ripening
(this is a thermally activated process in these systems, proceeding faster when bonds are weaker).  

The weakening of bonds at long times in the feedback process
arises due to properties of correlation and response functions during coarsening~\cite{crisanti03}.  Briefly, the coarsening of large
crystalline clusters depends on their macroscopic properties (curvature and surface tension), so applying the perturbation
$h_p$ does not affect whether the cluster containing particle $p$ should grow or shrink.  In this sense, coarsening is an
irreversible process.  As a result, the response $\chi$ is small
in the long-time (coarsening) regime, and the With-Feed protocol pushes the system to increasingly weak bonds. For very long times, it is possible that the With-Feed protocol may reduce $\eb$ so far that the bonds become too weak and the clusters disintegrate.  We did not observe this, although we do find a similar effect when the damping term $\gamma_i$ is not included in (\ref{equ:eb-update}).  These possibilities should be considered when applying this scheme to other systems.  For example, it may be useful in some systems to operate the feedback scheme only during early and intermediate stages of assembly,
and to revert to a `No-feed' method on very long times.  

In this discussion of Ostwald ripening, it is also worth recalling the neglect of hydrodynamic effects in the simulation model being used, and the use of single particle MC moves in implementing the dynamics.  Both these choices~\cite{spaeth11,whitelam11-molsim} act to reduce the diffusion of large clusters of particles.  This favours the Ostwald ripening mechanism for assembly, compared with the process by which diffusing clusters collide and fuse.  However, in evaluating the `With-feed' scheme presented here, we emphasise that we are comparing it with a `No-feed' scheme that uses the same model and simulation methods.  So the improvement obtained by allowing time-dependent interactions is significant for this particular model.  If the diffusion of large clusters were faster within the model (perhaps due to hydrodynamic effects), both the `No-feed' and the `With-feed' results would change.  However, assuming that the link between reversible bonding and effective assembly is maintained, we would expect the feedback scheme presented here to vary the interaction strength so as to facilitate effective self-assembly in that case too.  This hypothesis remains to be tested, as does the effectiveness of the feedback scheme in optimising assembly in other systems. 

\section{Outlook}

To conclude, we have introduced an automated method for tuning interaction parameters during self-assembly, so as to avoid kinetic trapping
and promote the formation of ordered structures. 
We have demonstrated in computer simulation that this method produces high-quality crystals, and we have emphasised 
that it can be generalised straightforwardly to other assembling systems.
There is also potential for applications in experiments: attractions between colloidal particles can now be manipulated in 
real-time~\cite{alsayed04,klajn07,hertlein08,elsner09,savage09,bonn09,taylor12}, and correlation-response measurements have also been 
made~\cite{bonn07a,oukris10,ruocco10}.  
We hope that the potential for automated self-assembly using real-time feedback will stimulate further studies in this area. 

One direction that might be promising in computer simulation is to consider feedback schemes that do not rely on correlation-response measurements, but use other observables to tune assembly protocols.
For example, one might use a structural measure within the feedback loop, to promote assembly.  It would be possible to might make interactions stronger when there are too few bonds in the system, and weaken interactions if there are too many non-crystalline bonds.  We have not investigated this approach so far, because it relies on the pathway to the ordered state being known \emph{a priori}: this is not the case in viral capsid assembly~\cite{hagan06,jack07} and there are many examples of non-trivial pathways in crystal nucleation too~\cite{frenkel97,whitelam2010}.  In those cases, the structural approach to optimising assembly becomes very hard to implement.  Instead, we have chosen to concentrate specifically on the kinetics of binding and unbinding events, aiming to avoid kinetic trapping, but leaving the system free to explore all pathways towards the assembled state.  A similar consideration applies to systems with several possible ordered states, such as crystal polymorphs.  The approach that we have described does not select for any specific polymorph: it allows the system to form whatever ordered state it prefers.  It would be interesting to explore polymorph selection within this kind of scheme, but we defer such issues to later publications.

\begin{acknowledgments}

We thank Mike Hagan, Steve Whitelam, Paddy Royall and Nigel Wilding for helpful discussions.
DK would also like to thank Sharon Glotzer for support and guidance. 
We are grateful to the Engineering and Physical Sciences Research Council (EPSRC) for funding through grants EP/G038074/1 and EP/I003797/1.
DK was supported in part by the U. S. Army Research Office under Grant Award No. W911NF-10-1-0518.

\begin{appendix}

\section{Implementation of feedback loop}

\subsection{Measurement of response functions}

In general, accurate estimation of the response function $\chi(t,t_i)$ requires good statistics, 
and our implementation of the feedback loop is designed to achieve this. 
We emphasise that while accurate estimation of $\chi$ does require some computational overhead, the usual alternative is to perform
many `No-feed' simulations over a range of $\eb$, in order to locate the good-assembly regime. 
That approach also requires significant CPU time: by using a feedback scheme, we focus the computational effort
on the parameter range in which assembly is most effective, and on the potential benefits of using time-dependent interactions in self-assembly.

We note that $\chi(t,t_i)$ is defined in terms of the response of a single representative particle.
To ensure that information from all particles is used as efficiently as possible to estimate $\chi(t,t_i)$, it is convenient 
to simulate a system where $h_p=+h$ for half of the particles in the system (chosen at random) and $h_p=-h$ for the other half~\cite{jack07,klotsa11}. 
This approach is discussed in detail in Ref.~\onlinecite{klotsa11}. 
To accurately estimate the linear response defined in (\ref{equ:chi}), we take $h=\tilde h T/\eb$
with $\tilde h=0.1$.  (Smaller values of $\tilde h$ lead to larger numerical uncertainty in $\chi$, while larger values of $\tilde h$ lead to
systematic errors because of non-linear response of the system at $O(\tilde h)^2$ and higher.)

In addition, to avoid statistical uncertainty in $\chi(t,t_i)$ arising from the specific choice of particles that have $h_p=\pm h$, we calculate
the average response in two steps.
At the beginning of each measurement stage, the configuration of the system
is saved and the measurement stage is simulated twice: the sets of particles with $h_p=\pm h$ are chosen at random for the first simulation while 
the second simulation uses fields $h_p$ that are opposite 
to the initial choice.  The response is averaged over the two simulations, ensuring that we correctly estimate $\chi(t_i,t_i)=0$, and 
minimizing errors for small $t-t_i$.

We also increase our statistical sampling by
simulating $M=8$ systems in parallel, each with $N=10,000$ particles.  The systems evolve independently during measurement
and relaxation stages, but the correlation and response functions in (\ref{equ:CC}) and (\ref{equ:chi}) are obtained by averaging
over the $M$ different simulations.  These combined measurements are then used to determine the duration of measurement and
relaxation stages and to obtain an updated value for $\eb$.  (The protocol for $\eb$ is therefore the same in each of the $M$ simulations.)

Finally, we 
note that if we measure $\tilde X>1$ then we take $\tilde X = 1$ in (\ref{equ:eb-update}).  We find that this occurs most often when bonds are weak,
in which case the true value is $\tilde X \approx 1$ but there is a large statistical uncertainty in our measurements of $\chi$ and hence $\tilde X$.
By estimating $\tilde X=1$ in this case, we minimise the sensitivity of our results to this uncertainty.

\subsection{Feedback parameters}

The parameters associated with the feedback scheme are $X_0$, $\delta X$, $\alpha$, $C^*$, and $X^*$.  
We emphasise that they are all dimensionless and our choices for their values rely only on the correlation between dynamical
reversibility and self-assembly.  In this sense, we expect suitable choices for these parameters to depend only weakly on the 
self-assembly process being investigated.
In this section we give a short discussion of the physical
considerations that control the choice of these parameters.

In general, the crucial features are (i) if the feedback loop is {\it too sensitive} to irreversible bonding then statistical noise in
measurements of reversibility may transiently reduce the bond strength between particles, leading to disintegration of the assembling
crystal; (ii) if the feedback loop is {\it not sensitive enough} to irreversible bonding then the system may aggregate into disordered clusters
so quickly that the feedback process is unable to respond, leading to disordered final states.  These competing requirements run through
the following discussion of the feedback parameters.

 We choose $X_0=0.8$ to be the target value for the normalised response $\tilde X$.  As discussed
in the text, we expect effective assembly when deviations from equilibrium are small but finite, consistent with a value
of $X_0$ that is fairly close to $1$.  Based on Refs.~\onlinecite{jack07,klotsa11,grant11pre}, we anticipate reasonable assembly at least for $0.7<X_0<0.9$.
If we choose a value closer to unity (say, $X_0=0.9$) then the feedback loop would 
produce more reversible behaviour, leading in general to weaker bonds.  We expect assembly to be slower in this case, but
the assembled product may be of higher quality.  Choosing a smaller target (say $X_0=0.7$) would promote
more rapid bond formation, possibly at the expense of more defects.  

We choose $\alpha=1$ as the sensitivity of the feedback loop to deviations from the target.  If $\alpha$ is larger then the system
is more sensitive to noise in the measurements of correlation and response.  On the other hand, if $\alpha$ is smaller then the system responds 
less quickly if $\eb$ is far from its optimal value.  Due to the presence of the damping factor $\gamma_i$, if the value of $\alpha$ is too large
then the system should correct for this as damping increases and the effective $\alpha$ falls.  We have taken $\alpha=1$ throughout for simplicity:
the scheme might be optimised either by varying $\alpha$ or indeed by investigating whether update rules other than (\ref{equ:eb-update}) lead
to more effective assembly.

We choose $\delta X=0.1$, which determines whether the damping factor $\gamma$ should increase between measurement stages.  
(The damping factor increases if $\tilde X$ is within $\delta X$ of its target.)  The inclusion of $\gamma$ is based on
the Robbins-Munro process in statistics~\cite{robbins51},  which indicates that convergence should be mostly independent of the scheme
used to vary $\gamma$ on time. (However, this conclusion is strictly valid only if the optimal extent of reversibility is independent of time).  
For the initial iterations of the feedback loop, we found that the behaviour depends very weakly on the time-dependence of $\gamma$ and
hence on $\delta X$.  At longer times, there is some dependence, although we do expect this to be quite weak.  (The case $\delta X=0$ involves
no damping in the update rule, which can result in very weak bonds and disintegration of growing crystallites.)

Finally the parameters $C^*=0.3$ and $X^*=0.6$ determine the duration of measurement stages.  Since the normalised correlation function $\tilde C(t,t_i)$
decays with time, from unity to near zero, the choice $C^*=0.3$ indicates that the system is reversible over a time scale that is comparable
to the typical lifetime of a bond in the system. 
Other choices for $C^*$ within the range $0.1-0.5$ correspond to a similar physical interpretation and
we would anticipate similar results.  (Of course, there is a broad distribution of bond lifetimes in the system, since 
bonds that are incorporated into assembled crystals may persist through the simulation.  However, the decay of $C(t,t_i)$
to $C^*$ indicates that a substantial amount of bonds have been made or broken, allowing a meaningful measurement of reversibilty.) Finally, the parameter $X^*$ is
included to quickly identify measurement stages in which kinetic trapping is taking place, and to avoid any further aggregation.  Again, the precise 
value of this parameter should not be too important, as long as it is significantly smaller than the target $X_0$ -- the larger $X^*$ is the
more quickly the system can respond to incipient kinetic trapping; while smaller values of $X_0$ ensure that the system is less sensitive
to noise in the measurements of $\chi$.

\end{appendix}

\end{acknowledgments}

\bibliographystyle{apsrev4-1}

\end{document}